Comment on "Atmospheric ionization by high-fluence, hard spectrum solar proton events and their probable appearance in the ice core archive" by *A.L. Melott et al.* [2016]

K.A. Duderstadt[1], J.E. Dibb[1], C.H. Jackman[2], C.E. Randall[3,4], N.A. Schwadron[1], S. C. Solomon[5], H.E. Spence[1]

1. Institute for the Study of Earth, Oceans, and Space, University of New Hampshire, USA

2. Emeritus, NASA Goddard Space Flight Center, Greenbelt, Maryland, USA

3. Laboratory for Atmospheric and Space Physics, University of Colorado, Boulder, Colorado, USA.

4. Department of Atmospheric and Oceanic Sciences, University of Colorado Boulder, Boulder, Colorado, USA.

5. High Altitude Observatory, National Center for Atmospheric Research, USA

Corresponding author: K.A. Duderstadt (duderstadtk@eos.sr.unh.edu)

**Key Points:**

- It is necessary to include pre-existing stratospheric $HNO_3$ in an analysis of nitrate deposition from SPEs.
- SPE-produced nitrate is negligible compared to the stratospheric background.

**Abstract**

*Melott et al.* [2016] suggest that individual solar proton events (SPEs) are detectable as nitrate ion spikes in ice cores. They use the high fluence, high energy ("hard spectrum") SPE of 23 February 1956 to calculate an enhancement of $HNO_3$ from the surface to 46 km that is equivalent to a ~120 ng cm$^{-2}$ nitrate ion spike observed in the GISP2H ice core. The *Melott et al.* [2016] approach is fundamentally flawed, since it considers only the absolute column burden of SPE-produced nitrate and not the pre-existing nitrate in the stratosphere. Modeling studies supported by extensive observations [*Duderstadt et al.*, 2014, 2016, and this comment] show background $HNO_3$ in the lower and middle stratosphere equivalent to 2000 to 3000 ng cm$^{-2}$ nitrate. These high levels of background nitrate must also be included when estimating SPE enhancements to the deposition of nitrate ions that might eventually be preserved in an ice core. The 1956 SPE results in less than a 5% increase in the column burden of atmospheric $HNO_3$, not large enough to explain the nitrate spike seen in the GISP2H ice core. Even extreme SPE enhancements cannot explain nitrate peaks (typically hundreds of percent increases) observed in the ice record [*Duderstadt et al.*, 2016]. Realistic mechanisms linking nitrate ions in ice cores to SPEs have not been established. It is time to move the search for indicators of SPEs away from nitrate ions: Nitrate ions cannot be used as proxies for individual SPEs in the ice core record.

## 1. Introduction

*Melott et al.* [2016] and *Duderstadt et al.* [2016] are listed as companion papers, however we did not see *Melott et al.* [2016] prior to publication. They both study the potential for solar proton events (SPEs) to enhance atmospheric nitric acid ($HNO_3$) deposition to account for observed nitrate ion spikes in ice cores but reach opposite conclusions. *Melott et al.* [2016] estimate the absolute amount of nitrate produced in the atmosphere by SPEs without considering the substantial pre-existing atmospheric column of nitric acid ($HNO_3$), much of which is in the lower stratosphere. *Duderstadt et al.* [2016] conduct comprehensive chemical-transport modeling that includes background $HNO_3$. This comment addresses these differences and also corrects misrepresentations regarding *Duderstadt et al.* [2016] as referenced by *Melott et al.* [2016] and *Sinnhuber* [2016].

## 2. Why pre-existing stratospheric nitric acid cannot be neglected

In Section 1.2, *Melott et al.* [2016] propose that a "fundamentally important difference [between the papers] is that [*Melott et al.*, 2016] examine the total amount of nitrate expected to be produced in the air column [by an SPE] and compare that with ice cores." We agree with this statement and believe that, by neglecting background $HNO_3$, *Melott et al.* [2016] incorrectly approach the question of whether an SPE can create a nitrate spike in polar snow that might be preserved and later observed in an ice core. In contrast, *Duderstadt et al.* [2016] assess the SPE source of nitrate relative to the atmospheric background, resolved vertically as well as in terms of column density. Accounting for pre-existing nitrate is essential because the chemical and dynamical processes capable of moving $HNO_3$ downward from the middle and lower stratosphere into the troposphere, where it is available for deposition to the surface, apply to all $HNO_3$ molecules regardless of source.

Much of the recent debate regarding the use of nitrate spikes as signatures of SPEs involves whether enhancements of $HNO_3$ produced through ionization in Earth's atmosphere by an SPE can be rapidly moved downward into the troposphere and then deposited onto the surface of the polar ice sheets in Greenland or Antarctica [e.g., *Smart et al.*, 2014; *Wolff et al.*, 2016; *Smart et al.*, 2016]. *Melott et al.* [2016] suggest that the combination of subsidence within the polar vortex and downward transport due to the formation of nitric acid trihydrate (NAT) particles in polar stratospheric clouds (PSCs) might be capable of moving all of the SPE-derived atmospheric $HNO_3$ between the surface and 45 km into surface snow within 1-2 months. *Smart et al.* [2014] suggest even shorter transport time scales, including for the 1956 event. But regardless of the rate at which $HNO_3$ can be transported downward, we emphasize that both subsidence and denitrification of the mid-stratosphere by downward transport of NAT apply to *all* of the $HNO_3$ that is present, including pre-existing $HNO_3$, with no way to selectively transport *only* the newly formed $HNO_3$ from the SPE.

Figure 1 shows vertical profiles of $HNO_3$ and $NO_y$ calculated by the Whole Atmosphere Community Climate Model (WACCM), averaged over the polar vortex in December and February. The figure also includes the profile of $HNO_3$ from the 23 Feb 1956 SPE, determined by repeating the methods outlined in *Melott et al.* [2016], using *Usoskin et al.* [2011] daily average ionization rates from CRAC:CRII integrated over one day to estimate $NO_x$ production, assuming immediate conversion to $HNO_3$. Background

HNO$_3$ in the Arctic lower stratosphere (10 km to 30 km) is almost 2 orders of magnitude more than *Melott et al.* [2016] calculate for the 1956 SPE. By integrating the HNO$_3$ profile from the 1956 SPE from the surface to ~45 km, *Melott et al.* [2016] calculate a cumulative column nitrate density of ~120 ng cm$^{-2}$, which matches the magnitude of the nitrate spike in the GISP2H core attributed to the 1956 SPE. Note that this spike is roughly a factor of 2 enhancement over the background ice core nitrate (Figure 4 of *Melott et al*. [2016]). In comparison, background WACCM nitrate column densities from 8 km to 45 km (not including minimal tropospheric contributions for reasons explained in Figure 1) are ~2400 ng cm$^{-2}$ for December and ~2800 ng cm$^{-2}$ for February. Consequently, an SPE like the 1956 event would increase the column burden of HNO$_3$ from 0 km to 45 km by only 5%. There is no plausible mechanism to get the HNO$_3$ SPE enhancement into the snow without also including pre-existing HNO$_3$. Thus any SPE-induced ice core enhancement could be no more than ~5%, much less than the 100% suggested by *Melott et al*. [2016].

The WACCM profiles in Figure 1 are model derived, but they clearly illustrate the reservoir of HNO$_3$ that results primarily from the oxidation of nitrous oxide (N$_2$O) emitted at the surface. Figure 2 compares WACCM profiles of HNO$_3$ mixing ratios with balloon and satellite measurements, showing that they substantially agree with observations and are accurate within, at worst, a factor of 2. *Brakebusch et al*. [2013] compare WACCM calculations to MLS observations, showing that WACCM HNO$_3$ in the lower stratosphere is 10% to 30% higher than MLS observations. Therefore, since even a very large SPE cannot significantly increase lower stratospheric HNO$_3$, we stand by our conclusions that SPEs cannot cause nitrates spikes in polar snow or ice cores.

**3. Correcting misrepresentations by *Melott et al.* [2016] and *Sinnhuber* [2016]**

There are two significant misrepresentations of our work. *Melott et al.* [2016] incorrectly define total $NO_y$ in our WACCM simulations, in the context of concerns about the contribution of tropospheric chemical species such as peroxyacetyl nitrate (PAN) and organic nitrates to total $NO_y$. In Section 1.2, *Melott et al.* [2016] erroneously claim that the "total $NO_y$ present in [the *Duderstadt et al.,* 2016] WACCM model atmospheric reservoir...includes many more species than they list in their definition." In *Duderstadt et al.* [2016] total odd nitrogen, $NO_y$, is defined as:

$$NO_y = N+NO+NO_2+NO_3+2N_2O_5+HNO_3+HO_2NO_2+ClONO_2+BrONO_2$$

(We note that nitric oxide, NO, was inadvertently omitted from this list of $NO_y$ species in the Introduction of *Duderstadt et al.* [2016] but is indeed a part of both the $NO_x$ and $NO_y$ families.)† The WACCM chemical mechanism used in *Duderstadt et al.* [2016] includes only these listed $NO_y$ species and their reactions and does not include more extensive organic tropospheric chemistry involving PAN and alkyl nitrates. WACCM is initialized using a previously completed climatological simulation and then integrated for four years to minimize the effects of tropospheric initial conditions. It is clear from Figure 1 that $NO_y$ is overwhelmingly composed of $HNO_3$ at the altitudes relevant to this discussion. Figure 2 shows how the WACCM predicted $HNO_3$ compares well with ILAS and MLS satellite measurements outside of SPE periods.

The *Melott et al.* [2016] assertion that there is "little to no ionization below 20 km" in our calculations is also inaccurate. *Sinnhuber* [2016] correctly points out that both studies use essentially equivalent methods of calculating atmospheric ionization and subsequent production of nitric oxides, but then incorrectly states, "ionization rates in the lowermost stratosphere and troposphere (below ~20 km) are higher in the *Melott et al.* [2016] scenario for 1956 than in any of the scenarios shown in *Duderstadt et al.* [2016], therefore leading to more direct production of nitrate there." Both studies use power law fits (similar to a Band function) to extrapolate to higher energies, *Usoskin et al.* [2010, 2011] CRAC:CRII yield functions rates to account for the effects of nuclear processes and secondary particles, and *Porter et al.* [1976] and *Rusch et al.* [1981] estimates of $NO_x$ production. Therefore, any major differences in ionization rates and $NO_x$ production should only depend on the solar proton flux at the top of the atmosphere.

---

† *Added 8 July 2016*

Figure 3 provides an example of ionization rates from *Duderstadt et al.* [2016] for 20 Jan 2005, adjusting contour levels to highlight ionization below 20 km. Ion pair production rates below 20 km are in the 100s $cm^{-3}$ $s^{-1}$, consistent with the observations of *Nicoll and Harrison* [2014] and calculations of *Melott et al.* [2016]. While the 20 Jan 2005 event in Figure 3 was short-lived, our suite of hypothetical events were both amplified and extended in length, representing effects of SPEs with significantly higher fluence and harder spectra than the 1956 SPE. The ionization rates, $NO_x$ production, and equivalent nitrate densities calculated during SPE events in *Duderstadt et al.* [2016] are therefore of the same magnitude or larger than in the *Melott et al.* [2016] study.

**4. Conclusions**

Using ice cores and other paleoarchives in conjunction with global climate models to interpret the historical behavior of the Sun remains exciting and promising, especially considering recent progress studying cosmogenic radionuclides [*Beer et al.,* 2012]. However, realistic mechanisms linking nitrate ions in ice cores to SPEs have not been established, and nitrate spikes in ice cores cannot provide statistically reliable proxy records of the frequency or magnitude of SPEs because of the many other causes of nitrate variability [e.g., *Legrand and Delmas,* 1986; *Wolff et al.,* 2008, 2012, 2016; *Duderstadt et al.,* 2014, 2016].

The arguments traditionally presented for associating SPEs with nitrate in ice cores rely on selectively choosing spikes that fall near dates of observed historical solar flares, such as the Carrington event of 1859 or neutron monitor enhancements in the 1940s and 1950s [e.g., *Zeller and Dreschoff,* 1995; *Kepko et al.,* 2009; *Smart et al.,* 2014; *Smart et al.,* 2016]. These selected spikes are then used to extrapolate other nitrate spikes to hypothetical solar storms.

*Wolff et al.* [2008, 2012, 2016] have convincingly associated nitrate spikes with biomass burning and other tropospheric sources in ice cores that provide a full suite of chemical measurements. In addition, they show that nitrate layers can be explained through the "fixing" of nitrate by sea salt or dust, post-depositional processes, and local meteorology. *Legrand and Delmas* [1987, 1988] and *Legrand and Kirchner* [1990] present results from ice core analyses and two-dimensional modeling studies that come to the same conclusion as our recent three-dimensional model simulations, namely that there are no known mechanisms to allow stratospheric enhancements of SPE-produced nitrate to be distinguishable from other sources of nitrate spikes at the surface. These alternative explanations are ignored, with a common theme that any contrary results are based on the wrong measurements at the wrong location at the wrong time [e.g., *Laird et al.,* 1998; *Smart et al.,* 2014, 2016; *Melott et al.,* 2016]. The *Melott et al.* [2016] calculations for the 1956 SPE are representative of the right measurements at the right location at the right time. However, even in this extreme case, the SPE enhancements of $HNO_3$ in the stratosphere have a negligible effect (less than 5%) on nitrate column burdens from the surface to 45 km. The hypothetical high fluence, hard spectra SPEs presented in *Duderstadt et al.* [2016], with larger ionization rates and $NO_x$ production in the lower stratosphere than the 1956 SPE, are also unable to explain nitrate peaks (typically hundreds of percent increases) observed in the ice core record.

*Melott et al.* [2016] study the possibility of producing nitrate peaks from SPEs to the exclusion of the atmospheric background. The authors study only the nitrate contributions that support their conclusion while neglecting all other material in the surrounding medium. It is time to move the search for indicators of solar activity away from nitrate ions: Nitrate ions cannot be used as proxies for individual SPEs in the ice core record. Existing and previous studies that utilize nitrate peaks in the ice core record to identify individual SPEs are flawed.

**Acknowledgements**

This work was supported by NSF grant 1135432 to the University of New Hampshire. The model data used to produce the analysis and figures for this study are available upon request from the corresponding author. We thank Ilya Usoskin for providing SPE-ionization rates from CRAC:CRII.

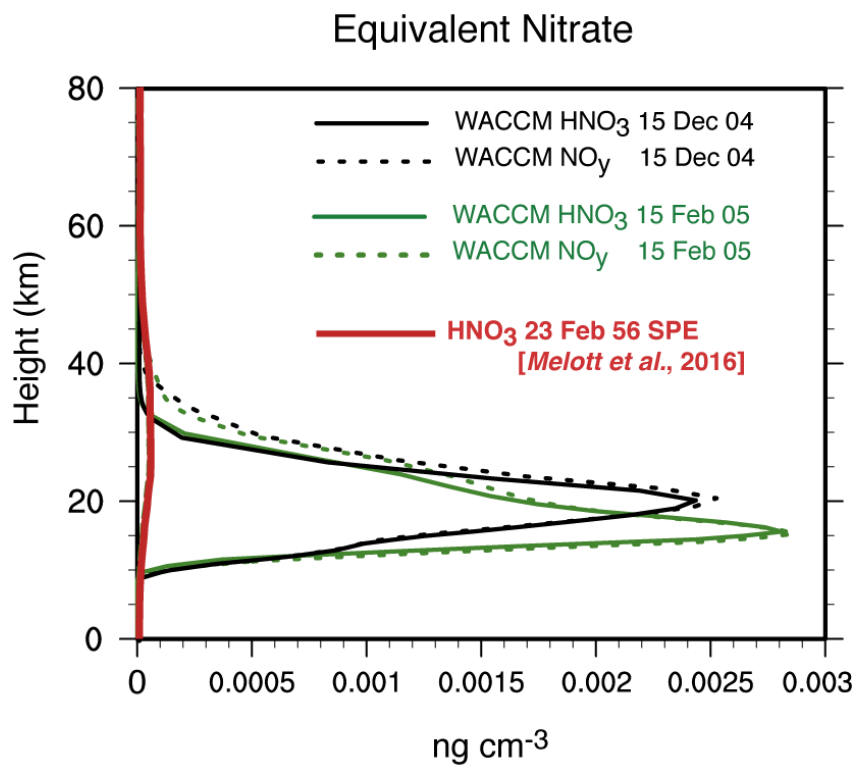

**Figure 1.** Vertical profiles of equivalent nitrate (NO$_3^-$) from background (no SPE) WACCM simulations from *Duderstadt et al.* [2016] and the 23 Feb 1956 SPE of *Melott et al.* [2016]. The profile of nitrate from HNO$_3$ for the 23 Feb 1956 SPE is based on the methods of *Melott et al.* [2016] using ionization rates from *Usoskin et al.* [2011] (red). WACCM profiles represent polar vortex averages of HNO$_3$ (solid) and total NO$_y$ (dashed) for December (black) and February (green). In order to emphasize the overwhelming contribution of background HNO$_3$ in the lower stratospheric to cumulative column nitrate, as well as to address concerns by *Melott et al.* [2015] over the contribution of polluted anthropogenic plumes to total column densities, we set WACCM values of HNO$_3$ and NO$_y$ below 8 km to zero in this figure. Although highly variable, HNO$_3$ in the WACCM polar troposphere generally ranges from $1\times10^{-4}$ to $3\times10^{-4}$ and NO$_y$ from $1\times10^{-4}$ to $5\times10^{-4}$, an order of magnitude less than in the lower stratosphere. This figure demonstrates that conclusions presented in *Duderstadt et al.* [2016] involving SPE enhancements to NO$_y$ column densities are independent of tropospheric NO$_y$ and primarily the result of this stratospheric reservoir.

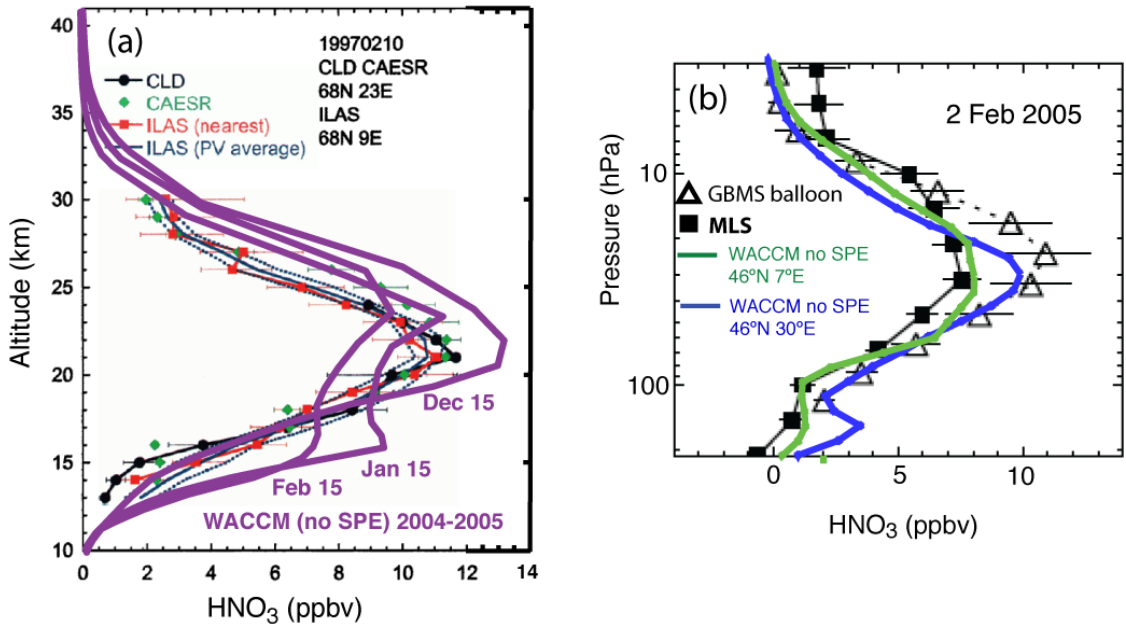

**Figure 2.** a) Vertical profiles of $HNO_3$ mixing ratios from WACCM (purple), balloon-borne measurements (black circles and green diamonds), and ILAS satellite measurements (blue solid lines with dashed lines as standard deviation). Adapted from *Koike et al.* [2000]. b) Vertical profiles of $HNO_3$ mixing ratios from WACCM (green and blue), balloon-borne measurements (triangles), and MLS satellite (black squares). MLS and balloon correspond to 46ºN 8ºE. All profiles are from 2 Feb 2005. Adapted from *Santee et al.* [2007].

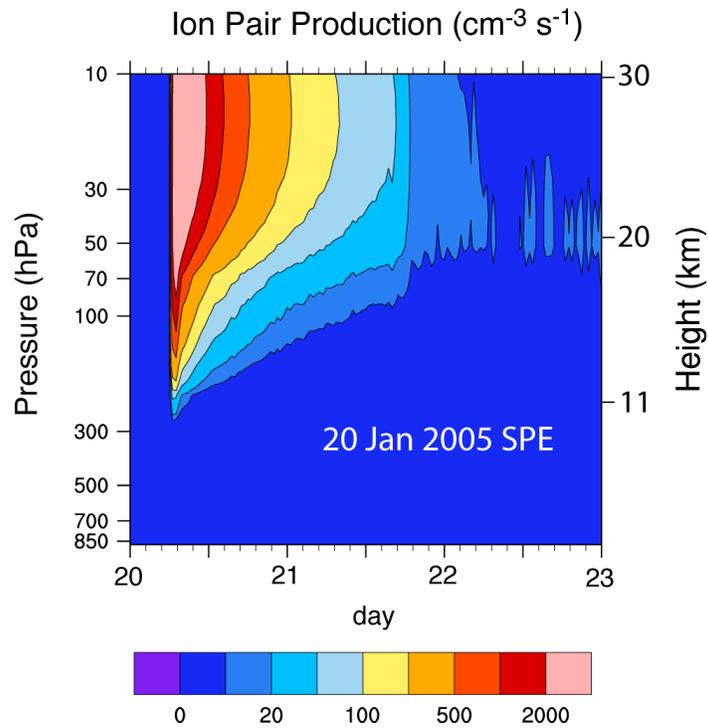

**Figure 3.** WACCM ion pair production rates ($cm^{-3}$ $s^{-1}$) for the 20 Jan 2005 SPE as a function of altitude (pressure and height) and time (day of the month).

17